\documentclass[prl,twocolumn,groupedaddress]{revtex4-1}

\usepackage{graphicx}
\usepackage{dcolumn}
\usepackage{bm}

\usepackage{amsmath}
\usepackage{graphicx}
\usepackage{color}
\usepackage{wrapfig}
\usepackage{epstopdf}

\graphicspath{{pics/}}

\newcommand{\bee}{\begin{equation}}
\newcommand{\ene}{\end{equation}}

\newcommand{\bea}{\begin{eqnarray}}
\newcommand{\ena}{\end{eqnarray}}

\begin{document}

\begin{figure}
\texttt{graphical abstract}
\includegraphics[width=0.4\textwidth]{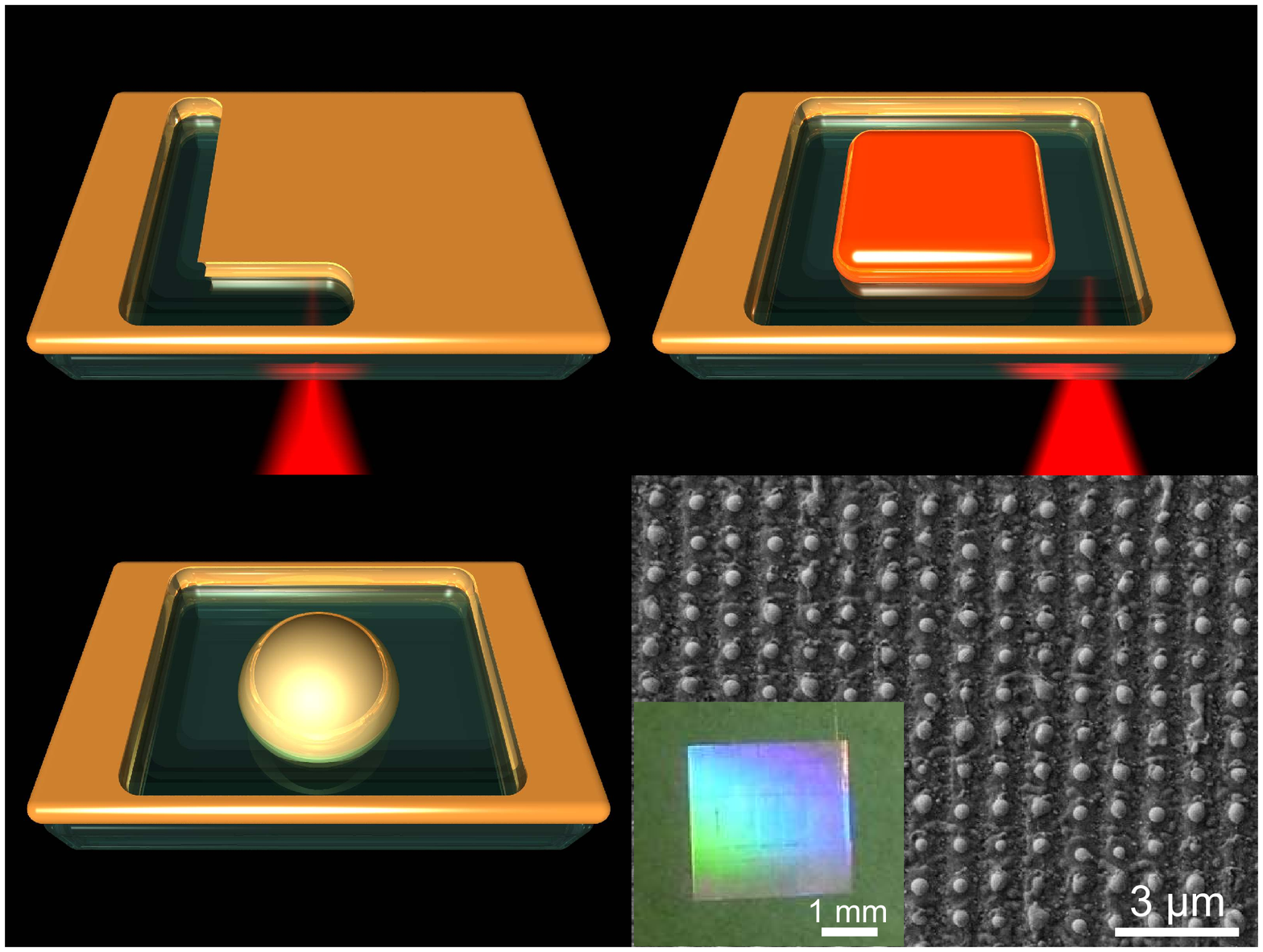}
\end{figure}

\title{Simple Method for Large-Scale Fabrication of Plasmonic Structures}

\author{Sergey V. Makarov$^{1}$, Valentin A. Milichko$^{1}$, Ivan S. Mukhin$^{1,2}$, Ivan I. Shishkin$^{1}$, \\ Alexey M. Mozharov$^{2}$, Alexander E. Krasnok$^{1}$, and Pavel A. Belov$^{1}$}

\address{$^{1}$Departament of Nanophotonics and Metamaterials, ITMO University, St.~Petersburg 197101, Russia\\
$^{2}$Laboratory of Renewable Energy Sources, St. Petersburg Academic University, St. Petersburg 194021, Russia}

\begin{abstract}
A novel method for single-step, lithography-free, and large-scale laser writing of nanoparticle-based plasmonic structures has been developed. Changing energy of femtosecond laser pulses and thickness of irradiated gold film it is possible to vary diameter of the gold nanoparticles, while the distance between them can be varied by laser scanning parameters. This method has an advantage over the most previously demonstrated methods in its simplicity and versatility, while the quality of the structures is good enough for many applications. In particular, resonant light absorbtion/scattering and surface-enhanced Raman scattering have been demonstrated on the fabricated nanostructures.

\end{abstract}

\maketitle

With the advent and rapid progress of fabrication technology in the past few decades, there has been growing interest in the field of plasmonics to explore novel phenomena by manipulating light at nanoscale~\cite{maier2007plasmonics, luk2010fano}. Examples of plasmonic devices requiring ordered nanoparticles are plasmonic filters~\cite{grigorenko08extremely, boltasseva2008plasmonic}, metasurfaces~\cite{meta2014capasso}, and waveguides~\cite{maier2003local}. The research area dealing with the interaction of molecules or molecular structures with plasmonic nanostructures is another rapidly growing field, due to potential analytical applications such as surface-enhanced Raman spectroscopy (SERS) and localized surface plasmon resonance spectroscopy~\cite{moskovits1985surface, Halas2007nano, anker2008biosensing}. In addition, plasmonic nanostructures exhibit strong nonlinear optical response that yields enhanced THz emission and high-harmonics generation~\cite{zayats2012nonlinear, alu2014giant, meta2015kivshar}.

To date, the most popular and controllable approaches of plasmonic nanostructures fabrication are based on direct ion-beam milling or multistage e-beam and nanoimprint lithographes. However, lithography-free and single-step methods are more desirable for large-scale manufacturing.~Among single-step techniques, chemical syntheses of monodisperse nanoparticles colloid is a promising method for high-throughput fabrication, however it requires additional technological steps to order the nanoparticles into functional nanostructure~\cite{Junno95,print2007nanoparticle,Shi2013NatCom, patra2014plasmofluidic}. An alternative cost-effective and versatile approach is to exploit self-assembly process via dewetting of heated thin supported film of a wide range of materials~\cite{Bischof96,dewetting2000anisotropic,Bobod2014wafer}. Laser radiation~\cite{Bischof96,Trice07,Wu11}, ion-beam~\cite{Lian06} or heating on a hotplate~\cite{dewetting2000anisotropic,Bobod2014wafer} have been utilized as heat sources for launching the dewetting process.

\begin{figure*}
\includegraphics[width=1\textwidth]{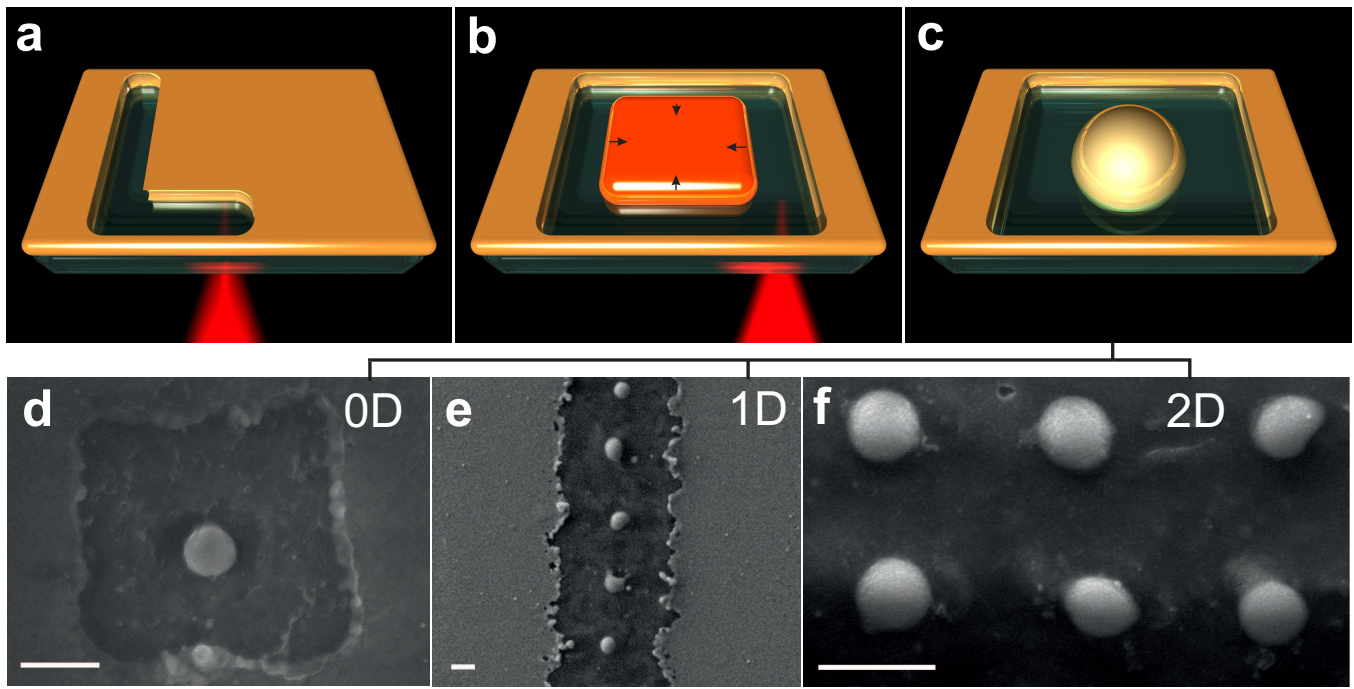}
\caption{A schematic illustrating of single nanoparticle formation under femtosecond laser cutting of patch from a gold film on a dielectric substrate (a-c). SEM images of a fabricated single gold nanoparticle (d), a line of the individual nanoparticles (e), and an array of of the individual nanoparticles (f) on a fused silica substrate fabricated by fs-laser irradiation of 30-nm Au film at fluence of 40 mJ/cm$^{\rm 2}$. Scale bars correspond to 500 nm.}\label{scheme}
\end{figure*}

Another way is to employ cost-effective laser technologies in order to controllably create nanoparticles via precise ablation of thin films, exploiting single nanoparticle transfer to a receiver substrate after single-shot femtosecond (fs) laser irradiation of a thin film~\cite{kuznetsov2009, kuznetsov2011, chichkov2014NatCom} or direct two-shots writing of plasmonic nanoantennas coupled with plasmonic lenses~\cite{MakarovOL2015}.

Despite many different approaches have been proposed to fabricate Au nanoparticles, the successful implementation of plasmonic devices still requires cost-effective and versatile method of controllable large-scale fabrication of ordered Au nanoparticles.

In this paper, we present a conceptually new, single-step, lithography-free and cost-efficient method for large-scale fabrication of nanoparticle-based plasmonic structures. The combination of simplicity of thin film dewetting process, along with precision and high-productivity of the laser technology enables extreme simplification of the process of nanoparticles fabrication, while the quality of fabricated structure is well-maintained. In other words, we achieve direct single-step ``writing" of nanopartices by strongly focused femtosecond laser pulses, where the laser irradiation is used to pattern and heat-up the residual film to the dewetting temperatures.~We implement this simple method for writing of 0D, 1D, and 2D plasmonic structures,  and show their applicability for the optical and sensing problems.
\begin{figure*}
\includegraphics[width=1\textwidth]{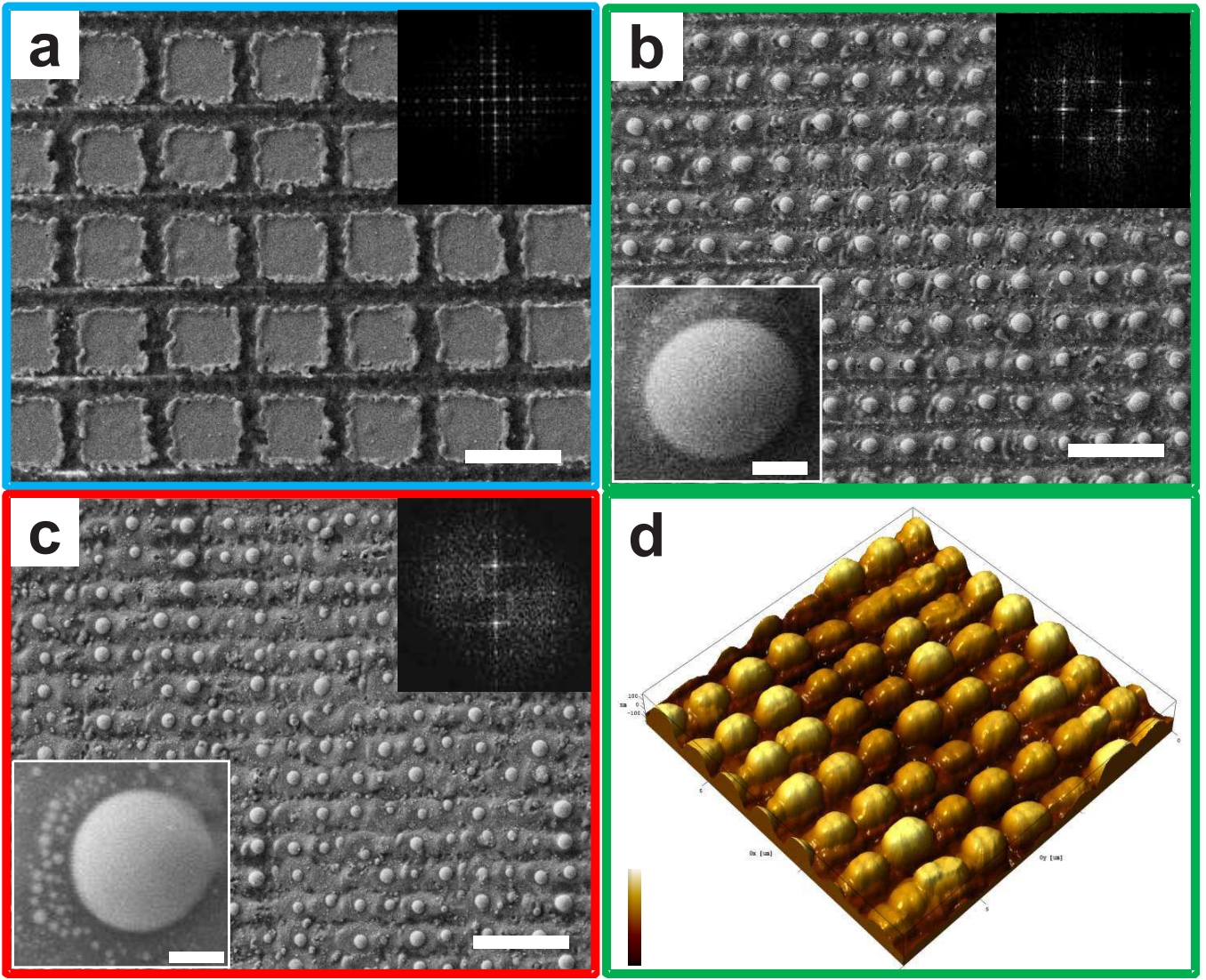}
\caption{SEM images of the Au nanoparticles arrays fabricated at fluence 40 mJ/cm$^{2}$ and periodicity of crossed laser scan of 2 (a), 1.0 (b), and 0.8 $\mu$m (c). Scale bars are 2 $\mu$m. The upper-right insets: Fourier spectra of the SEM images. The lower-left insets: enlarged SEM images of typical nanoparticles from the arrays with scale bars of 100 nm. (d) AFM image of the nanoparticles array, corresponding to the picture (b).}\label{2DSEM}
\end{figure*}

First, strongly focused fs-laser beam irradiates a thin Au film at fluence slightly higher than the thin film ablation threshold, which increases in range of 15--70 mJ/cm$^{2}$ with growth of the film thickness. These values are much lower than the values for single-shot ablation thresholds of thin gold films of corresponding thicknesses~\cite{kuznetsov2009, MakarovOL2015}. Since the scanning conditions correspond to the number of pulses \emph{N}$\sim$10$^{4}$ and the time delay between the fs-laser pulses in train is 12.5 ns, temperature in the vicinity of irradiated area is gradually accumulated pulse-by-pulse and achieves the level of film ablation in a certain region. Importantly, there is almost no ablation debris nearby each groove, being a result of evaporative mechanism of ablation, rather than spallative one followed by random dissemination of sub-100 nm nanoparticles in the vicinity of heated area~\cite{zhigilei2009ablation}. In this regime, rather clean and narrow (width of $\Delta$$\approx$400 nm$\leq$$\lambda$) grooves can be written directly on the surface (Fig.~\ref{scheme}). It is worth noting that femtosecond pulse duration minimizes the heat affected zone~\cite{bauerle2011laser}, resulting in a shallow groove depth of about 30\% deeper than the film thickness (see Supplementary Information).

Second, a single patch can be easily cut (Fig.~\ref{scheme}b) with any size larger than the uncertainty of the grooves edges ($\pm$50 nm, Fig.~\ref{scheme}d,e). The cut patch is thermally isolated since the thermal conductivities of the silica substrate and air are two and four orders of magnitude smaller than gold, respectively. Therefore, the isolated patch can be easier heated up to the temperatures where the film undergoes dewetting process.~The dewetting temperature of the films could be much lower than the melting point of bulk gold 1337 K. For instance, a 10-nm Au film on fused silica is dewetted even at 430 K, while 60-nm Au film undergoes dewetting at temperatures lower than 870 K~\cite{gadkari2005comparison}. Moreover, the smaller the patches (with smaller total volume), the higher the temperatures can be achieved at fixed fluence. As a result, the heated cut Au patch transforms into a nanoparticle of the same volume during the dewetting process (Fig.~\ref{scheme}).

Although dewetting is a spontaneous process for homogeneous film ~\cite{Bischof96, Trice07}, the reproducible formation of a certain number of nanoparticles has been demonstrated by heating of lithographicaly cut microscale patches of thin film~\cite{Lian06,Kim09APL,fowlkes2011self,wu2014directed,Bobod2014wafer}. Moreover, a single nanoparticle formation from each patch is possible, when patch width-to-height ratio ($\xi$=\emph{w/h}, where $w$ is the lateral size of the square patch and $h$ is the film thickness) is smaller than a certain value~\cite{Kim09APL}. In the case of Au film on SiO$_{\rm 2}$, this value must fulfill the condition $\xi$$<$40, to provide almost 100 percentage probability of single particle formation with a certain diameter~\cite{Kim09APL}. In our experiments we achieve the most controllable formation of nanoparticles at $\xi$$ \approx$15--30.

This technique enables direct writing of 0D (a single nanoparticle, Fig.~\ref{scheme}d), 1D (a line of nanoparticles, Fig.~\ref{scheme}e) and 2D (an array of nanoparticles, Fig.~\ref{scheme}f) nanoparticle-based structures. The 2D structures are produced by laser beam scanning of the film surface in two orthogonal directions, allowing extremely high production rate $\sim$1 mm$^{\rm 2}$/min at available speed of laser scanning $\sim$0.3 m/s. All types of the structures have been successfully fabricated on 20 and 30 nm Au films, while the other thicknesses (10 and 60 nm) provide much less controllable film patterning. Beside film thickness, another parameter that affects the diameter and quality of the nanoparticles is the width $w$ of the cut patch.

\begin{figure}[b]
\begin{center}
\includegraphics[width=0.43\textwidth]{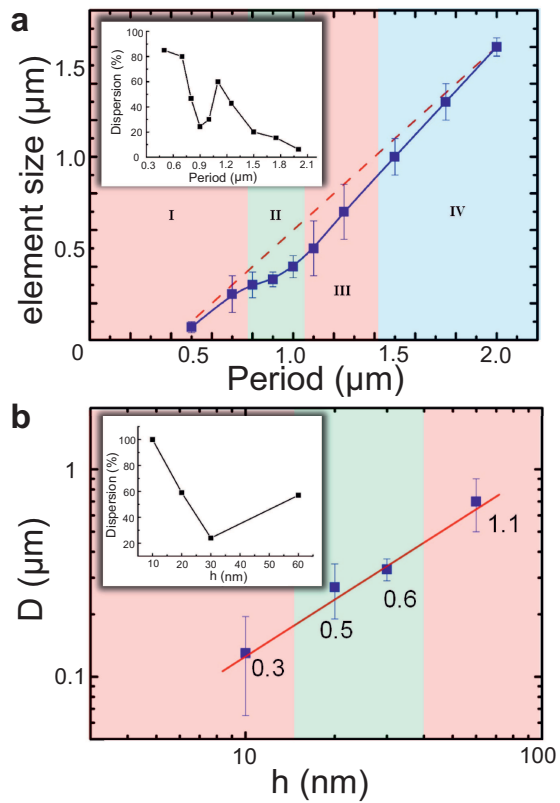}
\end{center}
\caption{(a) Measured dependence of a mean size of laser cut element on 30-nm Au film (blue squares) and the expected sizes calculated as the difference of the periods from groove width (red broken line). The inset represents dispersion of the size distribution as a function of the period of the crossed fs-laser scan. (b) Measured dependence of mean Au particle size on the film thickness at a given $w$. The inset represents dispersion of size distributions as a function of the film thickness. Red, green, and blue zones denote low, high, and moderate quality of the nanoparticles arrays, respectively, corresponding to colors of frames in Fig. 2.}\label{Dep}
\end{figure}

In Fig.~\ref{2DSEM}, SEM images of the fabricated 2D structures 30-nm Au film are shown as a sequence of reduced period between $90^{\circ}$-crossed laser scans at fluence of 40~mJ/cm$^{2}$. The laser fluence is chosen to provide as narrow and reproducible as possible grooves. Evidently, there are four main regimes: (i) square micro-patches formation (Fig.~\ref{2DSEM}a); (ii) unstable transition from square micro-patches to nanoparticles; (iii) formation of ordered and almost similar nanoparticles (Fig.~\ref{2DSEM}b,d); (iv) formation of disordered different-size nanoparticles (Fig.~\ref{2DSEM}c). The fabricated particles have almost spherical shape and are partly embedded into the substrate (Fig.~\ref{2DSEM}d).~Corresponding regimes are indicated in Fig.~\ref{Dep}a, where one can see that size-dispersion of cut patches and nanoparticles is a non-monotonic function of the scans period. In particular, for $h$=30 nm the size-dispersion demonstrates local minimum at scan period of 0.9--1 $\mu$m, i.e. in the most stable and reproducible regime of ordered nanoparticles formation for 30-nm Au film.

For film thickness of $h$=30~nm, we observe reproducible formation of sub-100 nm nanoparticles conglomerates nearby each ordered nanoparticle when periods of scan are smaller than 0.9 $\mu$m (see inset of Fig.~\ref{2DSEM}c). Similar behaviour has been observed when some parts of molten film are not absorbed by larger particles during dewetting process, and interpreted in terms of multimodal Rayleigh-Plateau instability~\cite{fowlkes2014hierarchical}. Indeed, the conglomerates are formed preferably from the side of last cutting of each patch (see left side in inset of Fig.~\ref{2DSEM}c).

The experimental thickness dependence of the nanoparticles diameters exhibits almost linear (slope is 0.9) dependence over the whole range of Au films thicknesses (Fig.~\ref{Dep}b), being a general trend for thin film dewetting ~\cite{thompson2012solid}. As it was mentioned before, we observe a transformation of each patch into a single nanoparticle, which is due to the too small sizes of the patches~\cite{Kim09APL} in comparison with typical wavelength of Rayleigh-type instability driven by capillary waves~\cite{landauFluid, Bischof96}. The value of wavelength of the dominating capillary wave on homogeneous film can be estimated as ~\cite{wyart1990drying}: $\lambda_{\rm m}=2\pi\sqrt{2/3} h^{2}/\alpha$, with the parameter $\alpha$ being characteristic of the material combination substrate/film ($\alpha$$\approx$4.3 nm for gold on fused silica~\cite{Bischof96}). Substituting our experimental parameters, we obtain the following estimations: $\lambda_{\rm m}$(h=10 nm)$\approx$0.12 $\mu$m, $\lambda_{\rm m}$($h$=20 nm)$\approx$0.48 $\mu$m, $\lambda_{\rm m}$($h$=30 nm)$\approx$1.1 $\mu$m, and $\lambda_{\rm m}$($h$=60 nm)$\approx$4.3 $\mu$m. Comparison of these values with corresponding values of $w$ plotted in Fig~\ref{Dep}b shows that the optimal conditions for single nanoparticle formation can not be fulfilled only for $h$=10~nm, which provides $\lambda_{\rm m}$$<$$w$ and comparable with the uncertainties ($\pm$ 50 nm) of grooves edges.

Similar to the scan period (the patch width) dependencies, there are also some range of thicknesses where quality of the 2D nanoparticles array is the most optimal (Fig~\ref{Dep}b). Non-monotonic behaviour of the size dispersion dependence on scan period and film thickness can be attributed to two main reasons: (i) the conditions $w/\lambda_{\rm m}$$>$1 and $\xi$$<$40 are not fulfilled for very small thicknesses and (ii) the dewetting temperature is much higher for thicker films. Therefore, too thin films can not provide ordered nanoparticles formation, while too thick and large patches can be hardly heated homogeneously during laser cutting to achieve controllable dewetting.

Gold nanoparticles are well known to provide resonant light absorption/scattering and local field enhancement in the optical range. In order to show high applicability of the proposed novel method, we examine basic optical properties of 2D nanoparticles arrays fabricated at different regimes: extinction and scattering spectra, and SERS.

\begin{figure}[!b]
\begin{center}
\includegraphics[width=0.47\textwidth]{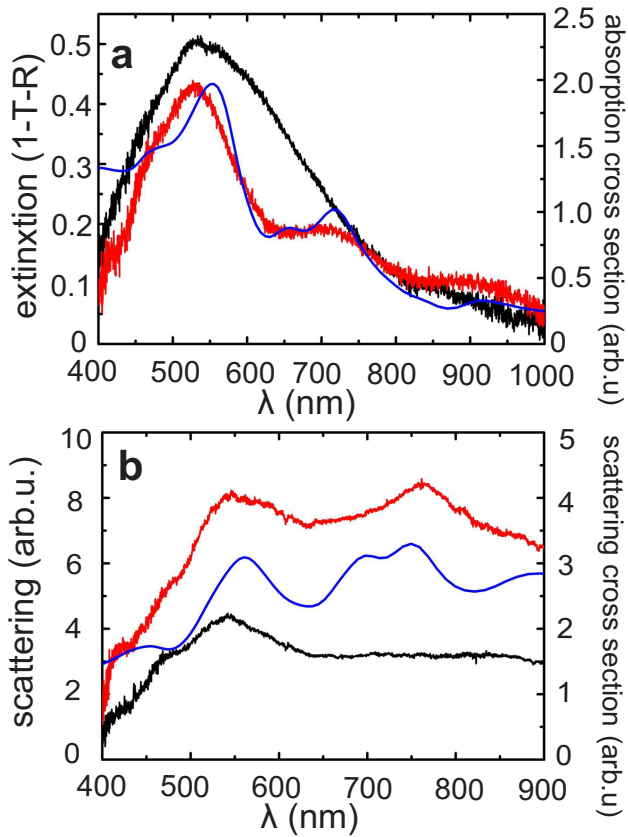}
\end{center}
\caption{Measured extinction (a) and scattering (b) spectra of ordered in 1$\times$1 $\mu$m  scanning regime (red curve, left axis) and disordered in 0.7$\times$0.7 $\mu$m scanning regime (black curve, left axis) gold nanoparticles. Numerically calculated absorbing (a) and scattering (b) cross-sections of a gold nanoparticles array (\textit{D} = 330 nm and \textit{P} = 1 $\mu$m) normalized on its geometrical cross-section (blue curve, right axis).}\label{Opt1}
\end{figure}

The measurements of extinction (Fig.~\ref{Opt1}a) and scattering (Fig.~\ref{Opt1}b) spectra from ordered 330-nm nanoparticles (Fig.~\ref{2DSEM}b,d) array revealed two resonances at $\lambda$$\approx$530 nm and $\lambda$$\approx$750 nm (for the measurements details see \textit{Methods}). These resonances are in good qualitative agreement with our numerical simulation (Fig.~\ref{Opt1}a,b) for array of gold 330-nm spheres embedded on half in fused silica infinite substrate (for calculation details see \textit{Methods}). According to our simulations, the "green" and "red" maxima correspond to quadrupole and dipole plasmon resonances, respectively. It is worth noting that in comparison with ordered monodisperse nanoparticles fabricated under optimal conditions (\textit{h}=30 nm, \textit{P}=1 $\mu$m), the disordered nanoparticles (\textit{h}=30 nm, \textit{P}=0.7 $\mu$m) with wide size distribution (from sub-100 nm up to 0.5 $\mu$m) exhibit single and broad extinction resonance. Indeed, sub-100 nm nanoparticles agglomerates surround each hundreds-nm particle (Fig.~\ref{2DSEM}c), providing additional strong dipole resonance contribution into scattering/absorbtion spectra in the range of $\lambda$=500--600 nm.

\begin{figure}[!t]
\begin{center}
\includegraphics[width=0.5\textwidth]{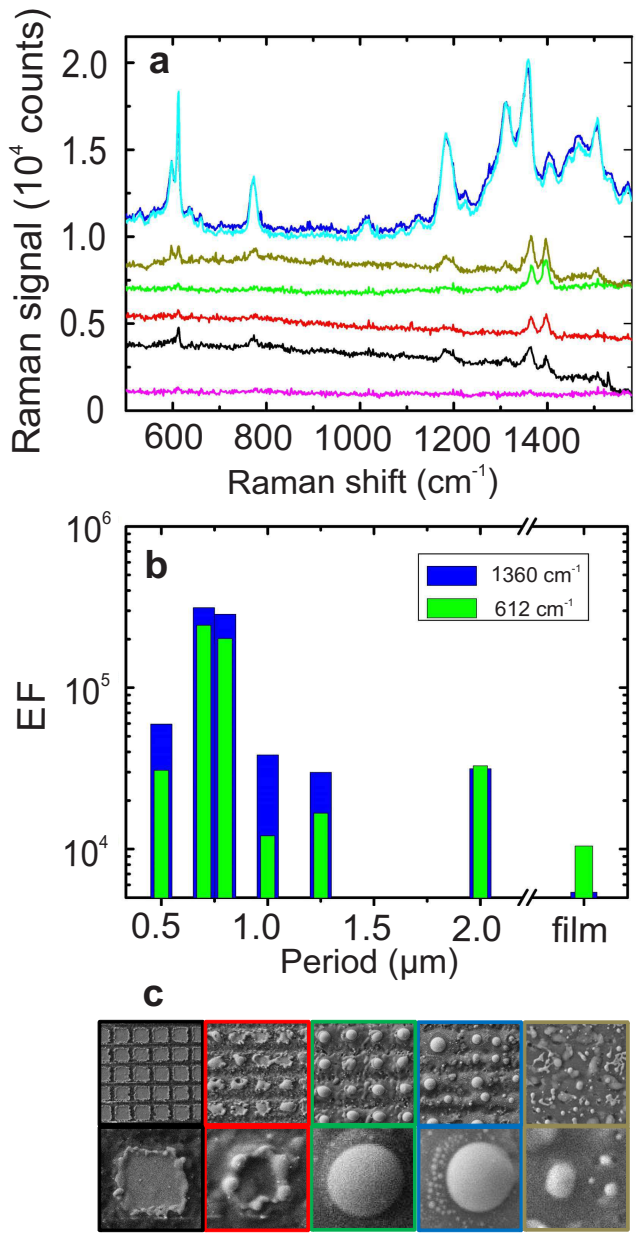}
\end{center}
\caption{(a) Raman scattering spectra from monolayer of R6G deposited on smooth (magenta curve) and patterned Au 30-nm film at fluence 40 mJ/cm$^{\rm 2}$ and different scan periods: 2$\mu$m (black), 1.25$\mu$m (red), 1$\mu$m (green), 0.8$\mu$m (deep blue), 0.7$\mu$m (blue), and 0.5$\mu$m (dark yellow).(b) Dependence of measured enhancement factor (averaged over a beam size) on period of fs-laser scanning during film pattering. (c) SEM-images of Au film surface patterned under the conditions, corresponding to the scan periods in picture (a) and having the same frames colors as Raman curves.}\label{Opt2}
\end{figure}

In order to show the applicability of the novel method of large-scale Au nanoparticles array fabrication, we measured Raman scattering signal from deposited molecules of rhodamine-6G (R6G) on the nanostructured substrates (for the measurements details, see \textit{Methods}). The typical Raman scattering spectra from a single layer of R6G on different fs-laser fabricated substrates with Au nanoparticles are shown in Fig.~\ref{Opt2}a. The systematic study of SERS signal on scan periods reveals its strong dependence on the type of nanostructures. The relatively low average enhancement factor (EF) $EF<10^{5}$ corresponds to both micro-patches and ordered nanoparticles fabricated at cross-scan periods \emph{P}$>$ 0.7$\mu$m, being approximately one order of magnitude higher than SERS signal from R6G on a smooth Au film (Fig.~\ref{Opt2}b). Nevertheless, these structures exhibit up to 5-fold higher $EF$ than a smooth Au 30-nm film with roughness amplitude less than 1 nm, which provides the so-called chemical mechanism of SERS and weak electromagnetic contribution on the roughness~\cite{moskovits1985surface}. The surface of the structures is not planar, and the main contribution to Raman signal gives R6G molecules from ordered Au nanoparticles with surface area of four times smaller than area of smooth substrate surface within laser spot. Taking into account this fact, the SERS signal for the ordered array of the nanoparticles is about one order of magnitude larger in comparison with the signal on the film. Our numerical modeling of the electric field distribution nearby a single 330-nm Au nanoparticle embedded in fused silica shows the highest $EF$ value lower than 100 in the hot spots, providing 3-fold electric field enhancement ($|E_{\rm loc}|/|E_{\rm inc}|$).

The strongest average enhancement factor $EF > 10^{\rm 5}$ is observed on structured 30-nm film at scan periods of 0.7 $\mu$m and 0.8 $\mu$m, where extremely hot spots are formed in the gaps within conglomerates of sub-100 nm nanoparticles (see the SEM images in deep blue frames in Fig.~\ref{Opt2}c). Our SEM images reveal only $\sim$10$^{\rm 2}$ pairs of the sub-100-nm nanoparticles separated on a few nanometers nearby each 330-nm Au nanoparticle Fig.~\ref{2DSEM}c,~\ref{Opt2}c). Therefore, one can assume that only $\sim$10$^{\rm 3}$--10$^{\rm 4}$ molecules from these gaps contribute to the average $EF$$>$10$^{\rm 5}$ over whole laser spot with $\approx$1.1$\times$10$^{6}$ molecules, resulting in local $EF$ in the gaps up to $\sim$10$^{\rm 9}$--10$^{\rm 10}$. Indeed, our numerical calculations of two touching gold 20-nm nanoparticles provide electric-field maximum enhancement at $\lambda$=633 nm in the range of $|E_{\rm loc}|/|E_{\rm inc}|$$\approx$10--20 in place of fitting of 1.5-nm R6G molecule between the particles, resulting in electric-field contribution to SERS local enhancement factor of $(|E_{\rm loc}|/|E_{\rm inc}|)^{4}$$\sim$10$^{5}$. Since the estimated chemical contribution to SERS from measurements of Raman signal of R6G on thin Au 30-nm film lies in the range of (5--9)$\times$10$^{\rm 3}$, the total \textit{EF} in the gap can achieve the level of $\sim$10$^{\rm 9}$, which is comparable with previous reports~\cite{pendry1996SERS, Halas2007gapSERS, ye2012plasmonic}.

In conclusion, the developed novel single-step and lithography-free method of direct Au-nanoparticles fabrication opens new possibilities to fabricate 0D, 1D, and 2D plasmonic nanostructures on large-scale. To the best of our knowledge, it is the first demonstration of simultaneous cutting and dewetting of thin metallic film processes that paves the way to extremely simplify the technology of monodisperse and ordered nanoparticles fabrication. The systematic experimental study reveals at least two functional regimes of film patterning: for light filtering and sensing. We believe that the applications range of this simple and low-cost method will be extended soon to nonlinear optical problems as well as various chemical and biomedical applications.

\textbf{Methods.} \textit{Fabrication.} A commercial femtosecond laser system (Femtosecond Oscillator TiF- 100F, Avesta Poject) was used, providing laser pulses at 780 nm central wavelength, with maximum pulse energy of 5 nJ, and pulse duration of 50 fs at the repetition rate of 80 MHz. Laser energy was varied and controlled by an acousto-optical modulator (R23080-3-LTD, Gooch and Housego) and a power meter (FielfMax II, Coherent), respectively, while the pulse duration was measured by an autocorrelator (Avesta Poject). Detailed experimental setup is presented in Supplementary Materials (Fig.S1).

Laser pulses were tightly focused by an oil immersion microscope objective (Carl Zeiss 100$\times$) with a numerical aperture (NA) of 1.4 (experimental setup is shown in Supplementary Information). According to the relation \emph{d}$\approx$1.22$\lambda$/NA, the estimated full-width at half-maximum diameter of the beam focal spot size is $d$=0.68~$\mu$m, which is close to measured value (0.64  $\mu$m) by standard method~\cite{liu1982simple} based on the dependence of laser-damaged area on incident laser energy (for details, see the Supplementary Information).

Laser beam was focused on supported Au films with thicknesses in the range of $h$=10--60 nm, thermally evaporated (Bock Edwards Auto 500) on the back side of 140-$\mu$m-thick silica glass without any additional adhesion-improving layers.~The samples were then placed on a three-dimensional air-bearing translating stage driven by brushless servomotors (ABL1000, Aerotech), allowing sample translation with various scan speeds up to 300 mm/s. In all experiments, film surface scanning by the laser beam was carried out with velocity of 5 mm/s, providing the number of laser pulses per each point of approximately $N\sim10^{4}$.

\textit{Characterizations}.  Preliminary optical imaging of the structured films was provided immediately during the laser processing by integrated CCD camera, collecting transmitted through the film white light from a lamp (Fig.S1). The high-resolution morphology characterization was carried out by means of scanning electron microscope (SEM, Carl Zeiss, Neon 40) and atomic force microscope (AFM, SmartSPM AIST-NT). The Fourier spectra of SEM images were obtained in ImajeJ software.

Optical transmission (\textit{T}) and reflection (\textit{R}) broadband measurements were carried out at normal incidence of linearly polarized light from a halogen lamp (HL-2000-FHSA), using a commercial spectrometer (Horiba LabRam HR) with CCD camera (Andor DU 420A-OE 325). The excitation Olympus PlanFI (NA=0.95) and collection Mitutoyo M Plan APO NIR (NA=0.7) objectives were used for transmission measurements. Objective Mitutoyo M Plan APO NIR (NA=0.42) was employed for reflection measurements. Scattering (\textit{S}) spectra measurements were carried out in a dark-field scheme, where the arrays irradiation was performed at an angle of incidence of 78$^{\circ}$ with surface normal and scattered signal collection was performed by means of objective Mitutoyo M Plan APO NIR (NA=0.7). Confocal optical scheme was optimized for collection of all signals (\textit{T},\textit{R} and \textit{S}).

\textit{SERS measurements and enhancement factor calculations.} The nanoparticles were functionalized with a selfassembled monolayer of R6G. The substrates were submerged in 0.5-mM solution made with water for 1 hour and then gently rinsed in neat water. Enhancement factor estimates were measured on a micro-Raman apparatus (Raman spectrometer HORIBA LabRam HR, AIST SmartSPM system). Using a 15-mW, 632.8-nm HeNe laser, spectra were recorded through the 50$\times$ microscope objective (NA=0.42) and projected onto a thermoelectrically cooled charge-coupled device (CCD, Andor DU 420A-OE 325) array using a 600-g/mm diffraction grating. Individual spectra were recorded from both single spots (1.8-$\mu$m diameter) on the substrates, and from a 1-cm thick cell of neat R6G for normalization.

From these measurements, the averaged over the excitation beam size Raman scattering enhancement factor (\emph{EF}) for the substrate is estimated with $EF_{\rm exp} = (I_{\rm SERS}/I_{\rm norm}) (N_{\rm norm}/N_{\rm SERS})$, where $I_{\rm SERS}$ and $I_{\rm norm}$ are the intensities of a specific Raman band, respectively. Two Raman modes of R6G at 612 cm$^{-1}$ and 1360 cm$^{-1}$ were chosen for the \emph{EF} calculations. The number of probed molecules in the unenhanced sample of neat R6G, $N_{\rm norm}$, is approximated using the molecular weight and density of R6G and the effective interaction volume of the Gaussian laser beam with the neat R6G sample. To estimate the $N_{\rm norm}$, we calculated the effective excitation volume by using the equation: $V_{\rm ex} = \pi r^{2}H$, where \textit{r} is the radius of the beam size ($r$$\approx$0.9 $\mu$m) and \emph{H} is the effective depth of focus ($H$$\approx$9 $\mu$m). Thus, we estimated an effective excitation volume of 22.9 $\mu m^{2}$ for our Raman microscopy with 633 nm excitation using the objective. Then, $N_{\rm norm}$ was calculated by the expression: $N_{\rm norm} = C N_{\rm A}V_{\rm ex}\rho/M = 1.08\times10^{9}$ molecules, where $C$ is the concentration ($3 \% $), $\rho$ is the density of R6G (1.26 g/cm$^{\rm 3}$), \textit{M} is the molar mass of R6G (479 g/mol) and $N_{\rm A}$ is the Avogadro constant (6.02$\times$10$^{\rm 23}$ mol$^{-1}$). To determine $N_{\rm SERS}$, a self-assembled monolayer of R6G molecules (molecular footprint size of S$_{\rm R6G}$$\approx$2.2 nm$^{\rm 2}$~\cite{gupta2003high}) was assumed to be closely packed on the surface and the number of the molecules within focused beam can be estimated as \emph{N}=$\pi$r$^{2}$/S$_{\rm R6G}$$\approx$1.1$\times$10$^{6}$.

\textit{Numerical Simulations.} Numerical simulations are performed by using both frequency-domain and time-domain solvers in the commercial software CST Microwave Studio. Incidence of a plane wave at given wavelength on a gold ~\cite{palik} sphere of a given radius embedded on half in fused silica ($\varepsilon$=2.1) infinite substrate is considered.

\textbf{Acknowledgments}

This work was financially supported by Russian Science Foundation (Grant 15-19-00172).
The authors are thankful to Prof. Y.S. Kivshar, Prof. S.I. Kudryashov, and M. Hasan for discussions.

\bibliographystyle{apsrev4-1}
\bibliography{Ablation_arxiv}

\begin{thebibliography}{42}%
\makeatletter
\providecommand \@ifxundefined [1]{%
 \@ifx{#1\undefined}
}%
\providecommand \@ifnum [1]{%
 \ifnum #1\expandafter \@firstoftwo
 \else \expandafter \@secondoftwo
 \fi
}%
\providecommand \@ifx [1]{%
 \ifx #1\expandafter \@firstoftwo
 \else \expandafter \@secondoftwo
 \fi
}%
\providecommand \natexlab [1]{#1}%
\providecommand \enquote  [1]{``#1''}%
\providecommand \bibnamefont  [1]{#1}%
\providecommand \bibfnamefont [1]{#1}%
\providecommand \citenamefont [1]{#1}%
\providecommand \href@noop [0]{\@secondoftwo}%
\providecommand \href [0]{\begingroup \@sanitize@url \@href}%
\providecommand \@href[1]{\@@startlink{#1}\@@href}%
\providecommand \@@href[1]{\endgroup#1\@@endlink}%
\providecommand \@sanitize@url [0]{\catcode `\\12\catcode `\$12\catcode
  `\&12\catcode `\#12\catcode `\^12\catcode `\_12\catcode `\%12\relax}%
\providecommand \@@startlink[1]{}%
\providecommand \@@endlink[0]{}%
\providecommand \url  [0]{\begingroup\@sanitize@url \@url }%
\providecommand \@url [1]{\endgroup\@href {#1}{\urlprefix }}%
\providecommand \urlprefix  [0]{URL }%
\providecommand \Eprint [0]{\href }%
\providecommand \doibase [0]{http://dx.doi.org/}%
\providecommand \selectlanguage [0]{\@gobble}%
\providecommand \bibinfo  [0]{\@secondoftwo}%
\providecommand \bibfield  [0]{\@secondoftwo}%
\providecommand \translation [1]{[#1]}%
\providecommand \BibitemOpen [0]{}%
\providecommand \bibitemStop [0]{}%
\providecommand \bibitemNoStop [0]{.\EOS\space}%
\providecommand \EOS [0]{\spacefactor3000\relax}%
\providecommand \BibitemShut  [1]{\csname bibitem#1\endcsname}%
\let\auto@bib@innerbib\@empty
\bibitem [{\citenamefont {Maier}(2007)}]{maier2007plasmonics}%
  \BibitemOpen
  \bibfield  {author} {\bibinfo {author} {\bibfnamefont {S.~A.}\ \bibnamefont
  {Maier}},\ }\href@noop {} {\emph {\bibinfo {title} {Plasmonics: fundamentals
  and applications}}}\ (\bibinfo  {publisher} {Springer Science : Business
  Media},\ \bibinfo {year} {2007})\BibitemShut {NoStop}%
\bibitem [{\citenamefont {Luk'yanchuk}\ \emph {et~al.}(2010)\citenamefont
  {Luk'yanchuk}, \citenamefont {Zheludev}, \citenamefont {Maier}, \citenamefont
  {Halas}, \citenamefont {Nordlander}, \citenamefont {Giessen},\ and\
  \citenamefont {Chong}}]{luk2010fano}%
  \BibitemOpen
  \bibfield  {author} {\bibinfo {author} {\bibfnamefont {B.}~\bibnamefont
  {Luk'yanchuk}}, \bibinfo {author} {\bibfnamefont {N.~I.}\ \bibnamefont
  {Zheludev}}, \bibinfo {author} {\bibfnamefont {S.~A.}\ \bibnamefont {Maier}},
  \bibinfo {author} {\bibfnamefont {N.~J.}\ \bibnamefont {Halas}}, \bibinfo
  {author} {\bibfnamefont {P.}~\bibnamefont {Nordlander}}, \bibinfo {author}
  {\bibfnamefont {H.}~\bibnamefont {Giessen}}, \ and\ \bibinfo {author}
  {\bibfnamefont {C.~T.}\ \bibnamefont {Chong}},\ }\href@noop {} {\bibfield
  {journal} {\bibinfo  {journal} {Nature materials}\ }\textbf {\bibinfo
  {volume} {9}},\ \bibinfo {pages} {707} (\bibinfo {year} {2010})}\BibitemShut
  {NoStop}%
\bibitem [{\citenamefont {Kravets}\ \emph {et~al.}(2008)\citenamefont
  {Kravets}, \citenamefont {Schedin},\ and\ \citenamefont
  {Grigorenko}}]{grigorenko08extremely}%
  \BibitemOpen
  \bibfield  {author} {\bibinfo {author} {\bibfnamefont {V.}~\bibnamefont
  {Kravets}}, \bibinfo {author} {\bibfnamefont {F.}~\bibnamefont {Schedin}}, \
  and\ \bibinfo {author} {\bibfnamefont {A.}~\bibnamefont {Grigorenko}},\
  }\href@noop {} {\bibfield  {journal} {\bibinfo  {journal} {Phys. Rev. Lett.}\
  }\textbf {\bibinfo {volume} {101}},\ \bibinfo {pages} {087403} (\bibinfo
  {year} {2008})}\BibitemShut {NoStop}%
\bibitem [{\citenamefont {Liu}\ \emph {et~al.}(2008)\citenamefont {Liu},
  \citenamefont {Boltasseva}, \citenamefont {Pedersen}, \citenamefont {Bakker},
  \citenamefont {Kildishev}, \citenamefont {Drachev},\ and\ \citenamefont
  {Shalaev}}]{boltasseva2008plasmonic}%
  \BibitemOpen
  \bibfield  {author} {\bibinfo {author} {\bibfnamefont {Z.}~\bibnamefont
  {Liu}}, \bibinfo {author} {\bibfnamefont {A.}~\bibnamefont {Boltasseva}},
  \bibinfo {author} {\bibfnamefont {R.~H.}\ \bibnamefont {Pedersen}}, \bibinfo
  {author} {\bibfnamefont {R.}~\bibnamefont {Bakker}}, \bibinfo {author}
  {\bibfnamefont {A.~V.}\ \bibnamefont {Kildishev}}, \bibinfo {author}
  {\bibfnamefont {V.~P.}\ \bibnamefont {Drachev}}, \ and\ \bibinfo {author}
  {\bibfnamefont {V.~M.}\ \bibnamefont {Shalaev}},\ }\href@noop {} {\bibfield
  {journal} {\bibinfo  {journal} {Metamaterials}\ }\textbf {\bibinfo {volume}
  {2}},\ \bibinfo {pages} {45} (\bibinfo {year} {2008})}\BibitemShut {NoStop}%
\bibitem [{\citenamefont {Yu}\ and\ \citenamefont
  {Capasso}(2014)}]{meta2014capasso}%
  \BibitemOpen
  \bibfield  {author} {\bibinfo {author} {\bibfnamefont {N.}~\bibnamefont
  {Yu}}\ and\ \bibinfo {author} {\bibfnamefont {F.}~\bibnamefont {Capasso}},\
  }\href@noop {} {\bibfield  {journal} {\bibinfo  {journal} {Nature Materials}\
  }\textbf {\bibinfo {volume} {13}},\ \bibinfo {pages} {139} (\bibinfo {year}
  {2014})}\BibitemShut {NoStop}%
\bibitem [{\citenamefont {Maier}\ \emph {et~al.}(2003)\citenamefont {Maier},
  \citenamefont {Kik}, \citenamefont {Atwater}, \citenamefont {Meltzer},
  \citenamefont {Harel}, \citenamefont {Koel},\ and\ \citenamefont
  {Requicha}}]{maier2003local}%
  \BibitemOpen
  \bibfield  {author} {\bibinfo {author} {\bibfnamefont {S.~A.}\ \bibnamefont
  {Maier}}, \bibinfo {author} {\bibfnamefont {P.~G.}\ \bibnamefont {Kik}},
  \bibinfo {author} {\bibfnamefont {H.~A.}\ \bibnamefont {Atwater}}, \bibinfo
  {author} {\bibfnamefont {S.}~\bibnamefont {Meltzer}}, \bibinfo {author}
  {\bibfnamefont {E.}~\bibnamefont {Harel}}, \bibinfo {author} {\bibfnamefont
  {B.~E.}\ \bibnamefont {Koel}}, \ and\ \bibinfo {author} {\bibfnamefont
  {A.~A.}\ \bibnamefont {Requicha}},\ }\href@noop {} {\bibfield  {journal}
  {\bibinfo  {journal} {Nature Materials}\ }\textbf {\bibinfo {volume} {2}},\
  \bibinfo {pages} {229} (\bibinfo {year} {2003})}\BibitemShut {NoStop}%
\bibitem [{\citenamefont {Moskovits}(1985)}]{moskovits1985surface}%
  \BibitemOpen
  \bibfield  {author} {\bibinfo {author} {\bibfnamefont {M.}~\bibnamefont
  {Moskovits}},\ }\href@noop {} {\bibfield  {journal} {\bibinfo  {journal}
  {Rev. Mod. Phys.}\ }\textbf {\bibinfo {volume} {57}},\ \bibinfo {pages} {783}
  (\bibinfo {year} {1985})}\BibitemShut {NoStop}%
\bibitem [{\citenamefont {Lal}\ \emph {et~al.}(2007)\citenamefont {Lal},
  \citenamefont {Link},\ and\ \citenamefont {Halas}}]{Halas2007nano}%
  \BibitemOpen
  \bibfield  {author} {\bibinfo {author} {\bibfnamefont {S.}~\bibnamefont
  {Lal}}, \bibinfo {author} {\bibfnamefont {S.}~\bibnamefont {Link}}, \ and\
  \bibinfo {author} {\bibfnamefont {N.~J.}\ \bibnamefont {Halas}},\ }\href@noop
  {} {\bibfield  {journal} {\bibinfo  {journal} {Nature Photonics}\ }\textbf
  {\bibinfo {volume} {1}},\ \bibinfo {pages} {641} (\bibinfo {year}
  {2007})}\BibitemShut {NoStop}%
\bibitem [{\citenamefont {Anker}\ \emph {et~al.}(2008)\citenamefont {Anker},
  \citenamefont {Hall}, \citenamefont {Lyandres}, \citenamefont {Shah},
  \citenamefont {Zhao},\ and\ \citenamefont {Van~Duyne}}]{anker2008biosensing}%
  \BibitemOpen
  \bibfield  {author} {\bibinfo {author} {\bibfnamefont {J.~N.}\ \bibnamefont
  {Anker}}, \bibinfo {author} {\bibfnamefont {W.~P.}\ \bibnamefont {Hall}},
  \bibinfo {author} {\bibfnamefont {O.}~\bibnamefont {Lyandres}}, \bibinfo
  {author} {\bibfnamefont {N.~C.}\ \bibnamefont {Shah}}, \bibinfo {author}
  {\bibfnamefont {J.}~\bibnamefont {Zhao}}, \ and\ \bibinfo {author}
  {\bibfnamefont {R.~P.}\ \bibnamefont {Van~Duyne}},\ }\href@noop {} {\bibfield
   {journal} {\bibinfo  {journal} {Nature Materials}\ }\textbf {\bibinfo
  {volume} {7}},\ \bibinfo {pages} {442} (\bibinfo {year} {2008})}\BibitemShut
  {NoStop}%
\bibitem [{\citenamefont {Kauranen}\ and\ \citenamefont
  {Zayats}(2012)}]{zayats2012nonlinear}%
  \BibitemOpen
  \bibfield  {author} {\bibinfo {author} {\bibfnamefont {M.}~\bibnamefont
  {Kauranen}}\ and\ \bibinfo {author} {\bibfnamefont {A.~V.}\ \bibnamefont
  {Zayats}},\ }\href@noop {} {\bibfield  {journal} {\bibinfo  {journal} {Nature
  Photonics}\ }\textbf {\bibinfo {volume} {6}},\ \bibinfo {pages} {737}
  (\bibinfo {year} {2012})}\BibitemShut {NoStop}%
\bibitem [{\citenamefont {Lee}\ \emph {et~al.}(2014)\citenamefont {Lee},
  \citenamefont {Tymchenko}, \citenamefont {Argyropoulos}, \citenamefont
  {Chen}, \citenamefont {Lu}, \citenamefont {Demmerle}, \citenamefont {Boehm},
  \citenamefont {Amann}, \citenamefont {Al{\`u}},\ and\ \citenamefont
  {Belkin}}]{alu2014giant}%
  \BibitemOpen
  \bibfield  {author} {\bibinfo {author} {\bibfnamefont {J.}~\bibnamefont
  {Lee}}, \bibinfo {author} {\bibfnamefont {M.}~\bibnamefont {Tymchenko}},
  \bibinfo {author} {\bibfnamefont {C.}~\bibnamefont {Argyropoulos}}, \bibinfo
  {author} {\bibfnamefont {P.-Y.}\ \bibnamefont {Chen}}, \bibinfo {author}
  {\bibfnamefont {F.}~\bibnamefont {Lu}}, \bibinfo {author} {\bibfnamefont
  {F.}~\bibnamefont {Demmerle}}, \bibinfo {author} {\bibfnamefont
  {G.}~\bibnamefont {Boehm}}, \bibinfo {author} {\bibfnamefont {M.-C.}\
  \bibnamefont {Amann}}, \bibinfo {author} {\bibfnamefont {A.}~\bibnamefont
  {Al{\`u}}}, \ and\ \bibinfo {author} {\bibfnamefont {M.~A.}\ \bibnamefont
  {Belkin}},\ }\href@noop {} {\bibfield  {journal} {\bibinfo  {journal}
  {Nature}\ }\textbf {\bibinfo {volume} {511}},\ \bibinfo {pages} {65}
  (\bibinfo {year} {2014})}\BibitemShut {NoStop}%
\bibitem [{\citenamefont {Minovich}\ \emph {et~al.}(2015)\citenamefont
  {Minovich}, \citenamefont {Miroshnichenko}, \citenamefont {Bykov},
  \citenamefont {Murzina}, \citenamefont {Neshev},\ and\ \citenamefont
  {Kivshar}}]{meta2015kivshar}%
  \BibitemOpen
  \bibfield  {author} {\bibinfo {author} {\bibfnamefont {A.~E.}\ \bibnamefont
  {Minovich}}, \bibinfo {author} {\bibfnamefont {A.~E.}\ \bibnamefont
  {Miroshnichenko}}, \bibinfo {author} {\bibfnamefont {A.~Y.}\ \bibnamefont
  {Bykov}}, \bibinfo {author} {\bibfnamefont {T.~V.}\ \bibnamefont {Murzina}},
  \bibinfo {author} {\bibfnamefont {D.~N.}\ \bibnamefont {Neshev}}, \ and\
  \bibinfo {author} {\bibfnamefont {Y.~S.}\ \bibnamefont {Kivshar}},\
  }\href@noop {} {\bibfield  {journal} {\bibinfo  {journal} {Laser Photonics
  Rev.}\ }\textbf {\bibinfo {volume} {9}},\ \bibinfo {pages} {195} (\bibinfo
  {year} {2015})}\BibitemShut {NoStop}%
\bibitem [{\citenamefont {Junno}\ \emph {et~al.}(1995)\citenamefont {Junno},
  \citenamefont {Deppert}, \citenamefont {Montelius},\ and\ \citenamefont
  {Samuelson}}]{Junno95}%
  \BibitemOpen
  \bibfield  {author} {\bibinfo {author} {\bibfnamefont {T.}~\bibnamefont
  {Junno}}, \bibinfo {author} {\bibfnamefont {K.}~\bibnamefont {Deppert}},
  \bibinfo {author} {\bibfnamefont {L.}~\bibnamefont {Montelius}}, \ and\
  \bibinfo {author} {\bibfnamefont {L.}~\bibnamefont {Samuelson}},\ }\href@noop
  {} {\bibfield  {journal} {\bibinfo  {journal} {Appl. Phys. Lett.}\ }\textbf
  {\bibinfo {volume} {66}},\ \bibinfo {pages} {3627} (\bibinfo {year}
  {1995})}\BibitemShut {NoStop}%
\bibitem [{\citenamefont {Kraus}\ \emph {et~al.}(2007)\citenamefont {Kraus},
  \citenamefont {Malaquin}, \citenamefont {Schmid}, \citenamefont {Riess},
  \citenamefont {Spencer},\ and\ \citenamefont {Wolf}}]{print2007nanoparticle}%
  \BibitemOpen
  \bibfield  {author} {\bibinfo {author} {\bibfnamefont {T.}~\bibnamefont
  {Kraus}}, \bibinfo {author} {\bibfnamefont {L.}~\bibnamefont {Malaquin}},
  \bibinfo {author} {\bibfnamefont {H.}~\bibnamefont {Schmid}}, \bibinfo
  {author} {\bibfnamefont {W.}~\bibnamefont {Riess}}, \bibinfo {author}
  {\bibfnamefont {N.~D.}\ \bibnamefont {Spencer}}, \ and\ \bibinfo {author}
  {\bibfnamefont {H.}~\bibnamefont {Wolf}},\ }\href@noop {} {\bibfield
  {journal} {\bibinfo  {journal} {Nature Nanotech.}\ }\textbf {\bibinfo
  {volume} {2}},\ \bibinfo {pages} {570} (\bibinfo {year} {2007})}\BibitemShut
  {NoStop}%
\bibitem [{\citenamefont {Shi}\ \emph {et~al.}(2013)\citenamefont {Shi},
  \citenamefont {Harris}, \citenamefont {Fenollosa}, \citenamefont {Rodriguez},
  \citenamefont {Korgel},\ and\ \citenamefont {Meseguer}}]{Shi2013NatCom}%
  \BibitemOpen
  \bibfield  {author} {\bibinfo {author} {\bibfnamefont {L.}~\bibnamefont
  {Shi}}, \bibinfo {author} {\bibfnamefont {J.~T.}\ \bibnamefont {Harris}},
  \bibinfo {author} {\bibfnamefont {R.}~\bibnamefont {Fenollosa}}, \bibinfo
  {author} {\bibfnamefont {X.}~\bibnamefont {Rodriguez}, \bibfnamefont
  {I.and~Lu}}, \bibinfo {author} {\bibfnamefont {B.~A.}\ \bibnamefont
  {Korgel}}, \ and\ \bibinfo {author} {\bibfnamefont {F.}~\bibnamefont
  {Meseguer}},\ }\href@noop {} {\bibfield  {journal} {\bibinfo  {journal}
  {Nature Comm.}\ } (\bibinfo {year} {2013})}\BibitemShut {NoStop}%
\bibitem [{\citenamefont {Patra}\ \emph {et~al.}(2014)\citenamefont {Patra},
  \citenamefont {Chikkaraddy}, \citenamefont {Tripathi}, \citenamefont
  {Dasgupta},\ and\ \citenamefont {Kumar}}]{patra2014plasmofluidic}%
  \BibitemOpen
  \bibfield  {author} {\bibinfo {author} {\bibfnamefont {P.~P.}\ \bibnamefont
  {Patra}}, \bibinfo {author} {\bibfnamefont {R.}~\bibnamefont {Chikkaraddy}},
  \bibinfo {author} {\bibfnamefont {R.~P.}\ \bibnamefont {Tripathi}}, \bibinfo
  {author} {\bibfnamefont {A.}~\bibnamefont {Dasgupta}}, \ and\ \bibinfo
  {author} {\bibfnamefont {G.~P.}\ \bibnamefont {Kumar}},\ }\href@noop {}
  {\bibfield  {journal} {\bibinfo  {journal} {Nature Comm.}\ }\textbf {\bibinfo
  {volume} {5}} (\bibinfo {year} {2014})}\BibitemShut {NoStop}%
\bibitem [{\citenamefont {Bischof}\ \emph {et~al.}(1996)\citenamefont
  {Bischof}, \citenamefont {Scherer}, \citenamefont {Herminghaus},\ and\
  \citenamefont {Leiderer}}]{Bischof96}%
  \BibitemOpen
  \bibfield  {author} {\bibinfo {author} {\bibfnamefont {J.}~\bibnamefont
  {Bischof}}, \bibinfo {author} {\bibfnamefont {D.}~\bibnamefont {Scherer}},
  \bibinfo {author} {\bibfnamefont {S.}~\bibnamefont {Herminghaus}}, \ and\
  \bibinfo {author} {\bibfnamefont {P.}~\bibnamefont {Leiderer}},\ }\href@noop
  {} {\bibfield  {journal} {\bibinfo  {journal} {Phys. Rev. Lett.}\ }\textbf
  {\bibinfo {volume} {77}},\ \bibinfo {pages} {1536} (\bibinfo {year}
  {1996})}\BibitemShut {NoStop}%
\bibitem [{\citenamefont {Higgins}\ and\ \citenamefont
  {Jones}(2000)}]{dewetting2000anisotropic}%
  \BibitemOpen
  \bibfield  {author} {\bibinfo {author} {\bibfnamefont {A.~M.}\ \bibnamefont
  {Higgins}}\ and\ \bibinfo {author} {\bibfnamefont {R.~A.}\ \bibnamefont
  {Jones}},\ }\href@noop {} {\bibfield  {journal} {\bibinfo  {journal}
  {Nature}\ }\textbf {\bibinfo {volume} {404}},\ \bibinfo {pages} {476}
  (\bibinfo {year} {2000})}\BibitemShut {NoStop}%
\bibitem [{\citenamefont {Abbarchi}\ \emph {et~al.}(2014)\citenamefont
  {Abbarchi}, \citenamefont {Naffouti}, \citenamefont {Vial}, \citenamefont
  {Benkouider}, \citenamefont {Lermusiaux}, \citenamefont {Favre},
  \citenamefont {Ronda}, \citenamefont {Bidault}, \citenamefont {Berbezier},\
  and\ \citenamefont {Bonod}}]{Bobod2014wafer}%
  \BibitemOpen
  \bibfield  {author} {\bibinfo {author} {\bibfnamefont {M.}~\bibnamefont
  {Abbarchi}}, \bibinfo {author} {\bibfnamefont {M.}~\bibnamefont {Naffouti}},
  \bibinfo {author} {\bibfnamefont {B.}~\bibnamefont {Vial}}, \bibinfo {author}
  {\bibfnamefont {A.}~\bibnamefont {Benkouider}}, \bibinfo {author}
  {\bibfnamefont {L.}~\bibnamefont {Lermusiaux}}, \bibinfo {author}
  {\bibfnamefont {L.}~\bibnamefont {Favre}}, \bibinfo {author} {\bibfnamefont
  {A.}~\bibnamefont {Ronda}}, \bibinfo {author} {\bibfnamefont
  {S.}~\bibnamefont {Bidault}}, \bibinfo {author} {\bibfnamefont
  {I.}~\bibnamefont {Berbezier}}, \ and\ \bibinfo {author} {\bibfnamefont
  {N.}~\bibnamefont {Bonod}},\ }\href@noop {} {\bibfield  {journal} {\bibinfo
  {journal} {ACS Nano}\ }\textbf {\bibinfo {volume} {8}},\ \bibinfo {pages}
  {11181} (\bibinfo {year} {2014})}\BibitemShut {NoStop}%
\bibitem [{\citenamefont {Trice}\ \emph {et~al.}(2007)\citenamefont {Trice},
  \citenamefont {Thomas}, \citenamefont {Favazza}, \citenamefont
  {Sureshkumar},\ and\ \citenamefont {Kalyanaraman}}]{Trice07}%
  \BibitemOpen
  \bibfield  {author} {\bibinfo {author} {\bibfnamefont {J.}~\bibnamefont
  {Trice}}, \bibinfo {author} {\bibfnamefont {D.}~\bibnamefont {Thomas}},
  \bibinfo {author} {\bibfnamefont {C.}~\bibnamefont {Favazza}}, \bibinfo
  {author} {\bibfnamefont {R.}~\bibnamefont {Sureshkumar}}, \ and\ \bibinfo
  {author} {\bibfnamefont {R.}~\bibnamefont {Kalyanaraman}},\ }\href@noop {}
  {\bibfield  {journal} {\bibinfo  {journal} {Phys. Rev. B}\ }\textbf {\bibinfo
  {volume} {75}},\ \bibinfo {pages} {235439} (\bibinfo {year}
  {2007})}\BibitemShut {NoStop}%
\bibitem [{\citenamefont {Wu}\ \emph {et~al.}(2011)\citenamefont {Wu},
  \citenamefont {Fowlkes},\ and\ \citenamefont {Rack}}]{Wu11}%
  \BibitemOpen
  \bibfield  {author} {\bibinfo {author} {\bibfnamefont {Y.}~\bibnamefont
  {Wu}}, \bibinfo {author} {\bibfnamefont {J.~D.}\ \bibnamefont {Fowlkes}}, \
  and\ \bibinfo {author} {\bibfnamefont {P.~D.}\ \bibnamefont {Rack}},\
  }\href@noop {} {\bibfield  {journal} {\bibinfo  {journal} {J. Mater. Res.}\
  }\textbf {\bibinfo {volume} {26}},\ \bibinfo {pages} {277} (\bibinfo {year}
  {2011})}\BibitemShut {NoStop}%
\bibitem [{\citenamefont {Lian}\ \emph {et~al.}(2006)\citenamefont {Lian},
  \citenamefont {Wang}, \citenamefont {Sun}, \citenamefont {Yu},\ and\
  \citenamefont {Ewing}}]{Lian06}%
  \BibitemOpen
  \bibfield  {author} {\bibinfo {author} {\bibfnamefont {J.}~\bibnamefont
  {Lian}}, \bibinfo {author} {\bibfnamefont {L.}~\bibnamefont {Wang}}, \bibinfo
  {author} {\bibfnamefont {X.}~\bibnamefont {Sun}}, \bibinfo {author}
  {\bibfnamefont {Q.}~\bibnamefont {Yu}}, \ and\ \bibinfo {author}
  {\bibfnamefont {R.~C.}\ \bibnamefont {Ewing}},\ }\href@noop {} {\bibfield
  {journal} {\bibinfo  {journal} {Nano Lett.}\ }\textbf {\bibinfo {volume} {6}}
  (\bibinfo {year} {2006})}\BibitemShut {NoStop}%
\bibitem [{\citenamefont {Kuznetsov}\ \emph {et~al.}(2009)\citenamefont
  {Kuznetsov}, \citenamefont {Koch},\ and\ \citenamefont
  {Chichkov}}]{kuznetsov2009}%
  \BibitemOpen
  \bibfield  {author} {\bibinfo {author} {\bibfnamefont {A.}~\bibnamefont
  {Kuznetsov}}, \bibinfo {author} {\bibfnamefont {J.}~\bibnamefont {Koch}}, \
  and\ \bibinfo {author} {\bibfnamefont {B.}~\bibnamefont {Chichkov}},\
  }\href@noop {} {\bibfield  {journal} {\bibinfo  {journal} {Opt. Express}\
  }\textbf {\bibinfo {volume} {17}},\ \bibinfo {pages} {18820} (\bibinfo {year}
  {2009})}\BibitemShut {NoStop}%
\bibitem [{\citenamefont {Kuznetsov}\ \emph {et~al.}(2011)\citenamefont
  {Kuznetsov}, \citenamefont {Evlyukhin}, \citenamefont {Gon{\c{c}}alves},
  \citenamefont {Reinhardt}, \citenamefont {Koroleva}, \citenamefont
  {Arnedillo}, \citenamefont {Kiyan}, \citenamefont {Marti},\ and\
  \citenamefont {Chichkov}}]{kuznetsov2011}%
  \BibitemOpen
  \bibfield  {author} {\bibinfo {author} {\bibfnamefont {A.~I.}\ \bibnamefont
  {Kuznetsov}}, \bibinfo {author} {\bibfnamefont {A.~B.}\ \bibnamefont
  {Evlyukhin}}, \bibinfo {author} {\bibfnamefont {M.~R.}\ \bibnamefont
  {Gon{\c{c}}alves}}, \bibinfo {author} {\bibfnamefont {C.}~\bibnamefont
  {Reinhardt}}, \bibinfo {author} {\bibfnamefont {A.}~\bibnamefont {Koroleva}},
  \bibinfo {author} {\bibfnamefont {M.~L.}\ \bibnamefont {Arnedillo}}, \bibinfo
  {author} {\bibfnamefont {R.}~\bibnamefont {Kiyan}}, \bibinfo {author}
  {\bibfnamefont {O.}~\bibnamefont {Marti}}, \ and\ \bibinfo {author}
  {\bibfnamefont {B.~N.}\ \bibnamefont {Chichkov}},\ }\href@noop {} {\bibfield
  {journal} {\bibinfo  {journal} {ACS Nano}\ }\textbf {\bibinfo {volume} {5}},\
  \bibinfo {pages} {4843} (\bibinfo {year} {2011})}\BibitemShut {NoStop}%
\bibitem [{\citenamefont {Zywietz}\ \emph {et~al.}(2014)\citenamefont
  {Zywietz}, \citenamefont {Evlyukhin}, \citenamefont {Reinhardt},\ and\
  \citenamefont {Chichkov}}]{chichkov2014NatCom}%
  \BibitemOpen
  \bibfield  {author} {\bibinfo {author} {\bibfnamefont {U.}~\bibnamefont
  {Zywietz}}, \bibinfo {author} {\bibfnamefont {A.~B.}\ \bibnamefont
  {Evlyukhin}}, \bibinfo {author} {\bibfnamefont {C.}~\bibnamefont
  {Reinhardt}}, \ and\ \bibinfo {author} {\bibfnamefont {B.~N.}\ \bibnamefont
  {Chichkov}},\ }\href@noop {} {\bibfield  {journal} {\bibinfo  {journal}
  {Nature Comm.}\ }\textbf {\bibinfo {volume} {5}} (\bibinfo {year}
  {2014})}\BibitemShut {NoStop}%
\bibitem [{\citenamefont {Ionin}\ \emph {et~al.}(2015)\citenamefont {Ionin},
  \citenamefont {Kudryashov}, \citenamefont {Makarov}, \citenamefont {Rudenko},
  \citenamefont {Kulchin}, \citenamefont {Vitrik}, \citenamefont {Efimov},\
  and\ \citenamefont {Kuchmizhak}}]{MakarovOL2015}%
  \BibitemOpen
  \bibfield  {author} {\bibinfo {author} {\bibfnamefont {A.}~\bibnamefont
  {Ionin}}, \bibinfo {author} {\bibfnamefont {S.}~\bibnamefont {Kudryashov}},
  \bibinfo {author} {\bibfnamefont {S.}~\bibnamefont {Makarov}}, \bibinfo
  {author} {\bibfnamefont {A.}~\bibnamefont {Rudenko}}, \bibinfo {author}
  {\bibfnamefont {Y.~N.}\ \bibnamefont {Kulchin}}, \bibinfo {author}
  {\bibfnamefont {O.}~\bibnamefont {Vitrik}}, \bibinfo {author} {\bibfnamefont
  {T.}~\bibnamefont {Efimov}}, \ and\ \bibinfo {author} {\bibfnamefont
  {A.}~\bibnamefont {Kuchmizhak}},\ }\href@noop {} {\bibfield  {journal}
  {\bibinfo  {journal} {Opt. Lett.}\ }\textbf {\bibinfo {volume} {40}},\
  \bibinfo {pages} {224} (\bibinfo {year} {2015})}\BibitemShut {NoStop}%
\bibitem [{\citenamefont {Zhigilei}\ \emph {et~al.}(2009)\citenamefont
  {Zhigilei}, \citenamefont {Lin},\ and\ \citenamefont
  {Ivanov}}]{zhigilei2009ablation}%
  \BibitemOpen
  \bibfield  {author} {\bibinfo {author} {\bibfnamefont {L.~V.}\ \bibnamefont
  {Zhigilei}}, \bibinfo {author} {\bibfnamefont {Z.}~\bibnamefont {Lin}}, \
  and\ \bibinfo {author} {\bibfnamefont {D.~S.}\ \bibnamefont {Ivanov}},\
  }\href@noop {} {\bibfield  {journal} {\bibinfo  {journal} {J. Phys. Chem. C}\
  }\textbf {\bibinfo {volume} {113}},\ \bibinfo {pages} {11892} (\bibinfo
  {year} {2009})}\BibitemShut {NoStop}%
\bibitem [{\citenamefont {Bauerle}(2011)}]{bauerle2011laser}%
  \BibitemOpen
  \bibfield  {author} {\bibinfo {author} {\bibfnamefont {D.~W.}\ \bibnamefont
  {Bauerle}},\ }\href@noop {} {\emph {\bibinfo {title} {Laser processing and
  chemistry}}}\ (\bibinfo  {publisher} {Springer Science \& Business Media},\
  \bibinfo {year} {2011})\BibitemShut {NoStop}%
\bibitem [{\citenamefont {Gadkari}\ \emph {et~al.}(2005)\citenamefont
  {Gadkari}, \citenamefont {Warren}, \citenamefont {Todi}, \citenamefont
  {Petrova},\ and\ \citenamefont {Coffey}}]{gadkari2005comparison}%
  \BibitemOpen
  \bibfield  {author} {\bibinfo {author} {\bibfnamefont {P.}~\bibnamefont
  {Gadkari}}, \bibinfo {author} {\bibfnamefont {A.}~\bibnamefont {Warren}},
  \bibinfo {author} {\bibfnamefont {R.}~\bibnamefont {Todi}}, \bibinfo {author}
  {\bibfnamefont {R.}~\bibnamefont {Petrova}}, \ and\ \bibinfo {author}
  {\bibfnamefont {K.}~\bibnamefont {Coffey}},\ }\href@noop {} {\bibfield
  {journal} {\bibinfo  {journal} {J. Vac. Sci. Tech. A}\ }\textbf {\bibinfo
  {volume} {23}},\ \bibinfo {pages} {1152} (\bibinfo {year}
  {2005})}\BibitemShut {NoStop}%
\bibitem [{\citenamefont {Kim}\ \emph {et~al.}(2009)\citenamefont {Kim},
  \citenamefont {Giermann},\ and\ \citenamefont {Thompson}}]{Kim09APL}%
  \BibitemOpen
  \bibfield  {author} {\bibinfo {author} {\bibfnamefont {D.}~\bibnamefont
  {Kim}}, \bibinfo {author} {\bibfnamefont {A.~L.}\ \bibnamefont {Giermann}}, \
  and\ \bibinfo {author} {\bibfnamefont {C.~V.}\ \bibnamefont {Thompson}},\
  }\href@noop {} {\bibfield  {journal} {\bibinfo  {journal} {Appl. Phys.
  Lett.}\ } (\bibinfo {year} {2009})}\BibitemShut {NoStop}%
\bibitem [{\citenamefont {Fowlkes}\ \emph {et~al.}(2011)\citenamefont
  {Fowlkes}, \citenamefont {Kondic}, \citenamefont {Diez}, \citenamefont {Wu},\
  and\ \citenamefont {Rack}}]{fowlkes2011self}%
  \BibitemOpen
  \bibfield  {author} {\bibinfo {author} {\bibfnamefont {J.~D.}\ \bibnamefont
  {Fowlkes}}, \bibinfo {author} {\bibfnamefont {L.}~\bibnamefont {Kondic}},
  \bibinfo {author} {\bibfnamefont {J.}~\bibnamefont {Diez}}, \bibinfo {author}
  {\bibfnamefont {Y.}~\bibnamefont {Wu}}, \ and\ \bibinfo {author}
  {\bibfnamefont {P.~D.}\ \bibnamefont {Rack}},\ }\href@noop {} {\bibfield
  {journal} {\bibinfo  {journal} {Nano Lett.}\ }\textbf {\bibinfo {volume}
  {11}},\ \bibinfo {pages} {2478} (\bibinfo {year} {2011})}\BibitemShut
  {NoStop}%
\bibitem [{\citenamefont {Wu}\ \emph {et~al.}(2014)\citenamefont {Wu},
  \citenamefont {Dong}, \citenamefont {Fu}, \citenamefont {Fowlkes},
  \citenamefont {Kondic}, \citenamefont {Vincenti}, \citenamefont {de~Ceglia},\
  and\ \citenamefont {Rack}}]{wu2014directed}%
  \BibitemOpen
  \bibfield  {author} {\bibinfo {author} {\bibfnamefont {Y.}~\bibnamefont
  {Wu}}, \bibinfo {author} {\bibfnamefont {N.}~\bibnamefont {Dong}}, \bibinfo
  {author} {\bibfnamefont {S.}~\bibnamefont {Fu}}, \bibinfo {author}
  {\bibfnamefont {J.~D.}\ \bibnamefont {Fowlkes}}, \bibinfo {author}
  {\bibfnamefont {L.}~\bibnamefont {Kondic}}, \bibinfo {author} {\bibfnamefont
  {M.~A.}\ \bibnamefont {Vincenti}}, \bibinfo {author} {\bibfnamefont
  {D.}~\bibnamefont {de~Ceglia}}, \ and\ \bibinfo {author} {\bibfnamefont
  {P.~D.}\ \bibnamefont {Rack}},\ }\href@noop {} {\bibfield  {journal}
  {\bibinfo  {journal} {ACS Appl. Mater. Interfaces}\ }\textbf {\bibinfo
  {volume} {6}},\ \bibinfo {pages} {5835} (\bibinfo {year} {2014})}\BibitemShut
  {NoStop}%
\bibitem [{\citenamefont {Fowlkes}\ \emph {et~al.}(2014)\citenamefont
  {Fowlkes}, \citenamefont {Roberts}, \citenamefont {Wu}, \citenamefont {Diez},
  \citenamefont {Gonzalez}, \citenamefont {Hartnett}, \citenamefont {Mahady},
  \citenamefont {Afkhami}, \citenamefont {Kondic},\ and\ \citenamefont
  {Rack}}]{fowlkes2014hierarchical}%
  \BibitemOpen
  \bibfield  {author} {\bibinfo {author} {\bibfnamefont {J.}~\bibnamefont
  {Fowlkes}}, \bibinfo {author} {\bibfnamefont {N.}~\bibnamefont {Roberts}},
  \bibinfo {author} {\bibfnamefont {Y.}~\bibnamefont {Wu}}, \bibinfo {author}
  {\bibfnamefont {J.}~\bibnamefont {Diez}}, \bibinfo {author} {\bibfnamefont
  {A.}~\bibnamefont {Gonzalez}}, \bibinfo {author} {\bibfnamefont
  {C.}~\bibnamefont {Hartnett}}, \bibinfo {author} {\bibfnamefont
  {K.}~\bibnamefont {Mahady}}, \bibinfo {author} {\bibfnamefont
  {S.}~\bibnamefont {Afkhami}}, \bibinfo {author} {\bibfnamefont
  {L.}~\bibnamefont {Kondic}}, \ and\ \bibinfo {author} {\bibfnamefont
  {P.}~\bibnamefont {Rack}},\ }\href@noop {} {\bibfield  {journal} {\bibinfo
  {journal} {Nano Lett.}\ }\textbf {\bibinfo {volume} {14}},\ \bibinfo {pages}
  {774} (\bibinfo {year} {2014})}\BibitemShut {NoStop}%
\bibitem [{\citenamefont {Thompson}(2012)}]{thompson2012solid}%
  \BibitemOpen
  \bibfield  {author} {\bibinfo {author} {\bibfnamefont {C.~V.}\ \bibnamefont
  {Thompson}},\ }\href@noop {} {\bibfield  {journal} {\bibinfo  {journal}
  {Annu. Rev. Mater. Res.}\ }\textbf {\bibinfo {volume} {42}},\ \bibinfo
  {pages} {399} (\bibinfo {year} {2012})}\BibitemShut {NoStop}%
\bibitem [{\citenamefont {Landau}\ and\ \citenamefont
  {Lifshitz}(1987)}]{landauFluid}%
  \BibitemOpen
  \bibfield  {author} {\bibinfo {author} {\bibfnamefont {L.}~\bibnamefont
  {Landau}}\ and\ \bibinfo {author} {\bibfnamefont {E.}~\bibnamefont
  {Lifshitz}},\ }\href
  {http://www.amazon.com/Fluid-Mechanics-Second-Edition-Theoretical/dp/0750627670}
  {\emph {\bibinfo {title} {Fluid Mechanics}}},\ \bibinfo {edition} {2nd}\ ed.\
  (\bibinfo  {publisher} {Butterworth-Heinemann},\ \bibinfo {year} {1987})\ p.\
  \bibinfo {pages} {551}\BibitemShut {NoStop}%
\bibitem [{\citenamefont {Wyart}\ and\ \citenamefont
  {Daillant}(1990)}]{wyart1990drying}%
  \BibitemOpen
  \bibfield  {author} {\bibinfo {author} {\bibfnamefont {F.~B.}\ \bibnamefont
  {Wyart}}\ and\ \bibinfo {author} {\bibfnamefont {J.}~\bibnamefont
  {Daillant}},\ }\href@noop {} {\bibfield  {journal} {\bibinfo  {journal} {Can.
  J. Phys.}\ }\textbf {\bibinfo {volume} {68}},\ \bibinfo {pages} {1084}
  (\bibinfo {year} {1990})}\BibitemShut {NoStop}%
\bibitem [{\citenamefont {Garc{\'\i}a-Vidal}\ and\ \citenamefont
  {Pendry}(1996)}]{pendry1996SERS}%
  \BibitemOpen
  \bibfield  {author} {\bibinfo {author} {\bibfnamefont {F.}~\bibnamefont
  {Garc{\'\i}a-Vidal}}\ and\ \bibinfo {author} {\bibfnamefont {J.}~\bibnamefont
  {Pendry}},\ }\href@noop {} {\bibfield  {journal} {\bibinfo  {journal} {Phys.
  Rev. Lett.}\ }\textbf {\bibinfo {volume} {77}},\ \bibinfo {pages} {1163}
  (\bibinfo {year} {1996})}\BibitemShut {NoStop}%
\bibitem [{\citenamefont {Ward}\ \emph {et~al.}(2007)\citenamefont {Ward},
  \citenamefont {Grady}, \citenamefont {Levin}, \citenamefont {Halas},
  \citenamefont {Wu}, \citenamefont {Nordlander},\ and\ \citenamefont
  {Natelson}}]{Halas2007gapSERS}%
  \BibitemOpen
  \bibfield  {author} {\bibinfo {author} {\bibfnamefont {D.~R.}\ \bibnamefont
  {Ward}}, \bibinfo {author} {\bibfnamefont {N.~K.}\ \bibnamefont {Grady}},
  \bibinfo {author} {\bibfnamefont {C.~S.}\ \bibnamefont {Levin}}, \bibinfo
  {author} {\bibfnamefont {N.~J.}\ \bibnamefont {Halas}}, \bibinfo {author}
  {\bibfnamefont {Y.}~\bibnamefont {Wu}}, \bibinfo {author} {\bibfnamefont
  {P.}~\bibnamefont {Nordlander}}, \ and\ \bibinfo {author} {\bibfnamefont
  {D.}~\bibnamefont {Natelson}},\ }\href@noop {} {\bibfield  {journal}
  {\bibinfo  {journal} {Nano Lett.}\ }\textbf {\bibinfo {volume} {7}},\
  \bibinfo {pages} {1396} (\bibinfo {year} {2007})}\BibitemShut {NoStop}%
\bibitem [{\citenamefont {Ye}\ \emph {et~al.}(2012)\citenamefont {Ye},
  \citenamefont {Wen}, \citenamefont {Sobhani}, \citenamefont {Lassiter},
  \citenamefont {Dorpe}, \citenamefont {Nordlander},\ and\ \citenamefont
  {Halas}}]{ye2012plasmonic}%
  \BibitemOpen
  \bibfield  {author} {\bibinfo {author} {\bibfnamefont {J.}~\bibnamefont
  {Ye}}, \bibinfo {author} {\bibfnamefont {F.}~\bibnamefont {Wen}}, \bibinfo
  {author} {\bibfnamefont {H.}~\bibnamefont {Sobhani}}, \bibinfo {author}
  {\bibfnamefont {J.~B.}\ \bibnamefont {Lassiter}}, \bibinfo {author}
  {\bibfnamefont {P.~V.}\ \bibnamefont {Dorpe}}, \bibinfo {author}
  {\bibfnamefont {P.}~\bibnamefont {Nordlander}}, \ and\ \bibinfo {author}
  {\bibfnamefont {N.~J.}\ \bibnamefont {Halas}},\ }\href@noop {} {\bibfield
  {journal} {\bibinfo  {journal} {Nano Lett.}\ }\textbf {\bibinfo {volume}
  {12}},\ \bibinfo {pages} {1660} (\bibinfo {year} {2012})}\BibitemShut
  {NoStop}%
\bibitem [{\citenamefont {Liu}(1982)}]{liu1982simple}%
  \BibitemOpen
  \bibfield  {author} {\bibinfo {author} {\bibfnamefont {J.}~\bibnamefont
  {Liu}},\ }\href@noop {} {\bibfield  {journal} {\bibinfo  {journal} {Opt.
  Lett.}\ }\textbf {\bibinfo {volume} {7}},\ \bibinfo {pages} {196} (\bibinfo
  {year} {1982})}\BibitemShut {NoStop}%
\bibitem [{\citenamefont {Gupta}\ and\ \citenamefont
  {Weimer}(2003)}]{gupta2003high}%
  \BibitemOpen
  \bibfield  {author} {\bibinfo {author} {\bibfnamefont {R.}~\bibnamefont
  {Gupta}}\ and\ \bibinfo {author} {\bibfnamefont {W.}~\bibnamefont {Weimer}},\
  }\href@noop {} {\bibfield  {journal} {\bibinfo  {journal} {Chem. Phys.
  Lett.}\ }\textbf {\bibinfo {volume} {374}},\ \bibinfo {pages} {302} (\bibinfo
  {year} {2003})}\BibitemShut {NoStop}%
\bibitem [{\citenamefont {Palik}(1998)}]{palik}%
  \BibitemOpen
  \bibfield  {author} {\bibinfo {author} {\bibfnamefont {E.~D.}\ \bibnamefont
  {Palik}},\ }\href@noop {} {\emph {\bibinfo {title} {Handbook of optical
  constants of solids}}}\ (\bibinfo  {publisher} {Academic press},\ \bibinfo
  {year} {1998})\BibitemShut {NoStop}%
\end{thebibliography}%

\newpage

\textbf{Supplementary materialss:\\ Simple Method for Large-Scale Fabrication of Plasmonic Structures}

\begin{figure*}
\begin{center}
\includegraphics[width=0.6\textwidth]{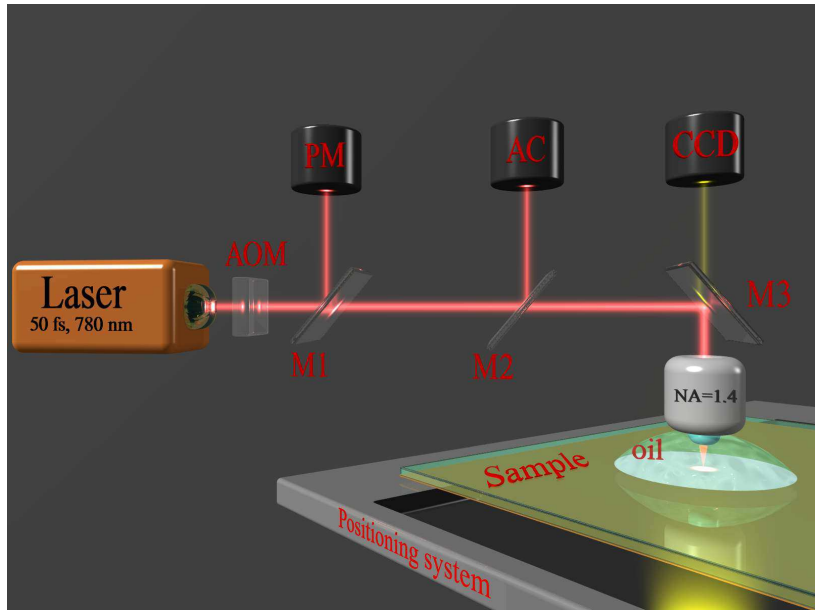}
\end{center}
\caption{Schematic illustration of experimental setup for laser ablation, where AOM is the acousto-optical modulator; M1 and M2 are the 100\%-dielectric mirrors (at wavelength 780 nm) on flipping mount; M3 is the fixed 100\% dielectric mirror, reflecting the laser beam to the oil immersion objective ($NA$=1.4); PM is the power meter; AC is the autocorrelator; CCD is the CCD-camera}\label{schemeS1}
\end{figure*}

Our experimental setup for laser ablation is schematically illustrated in Fig.~\ref{schemeS1}.

\begin{figure*}
\includegraphics[width=0.6\textwidth]{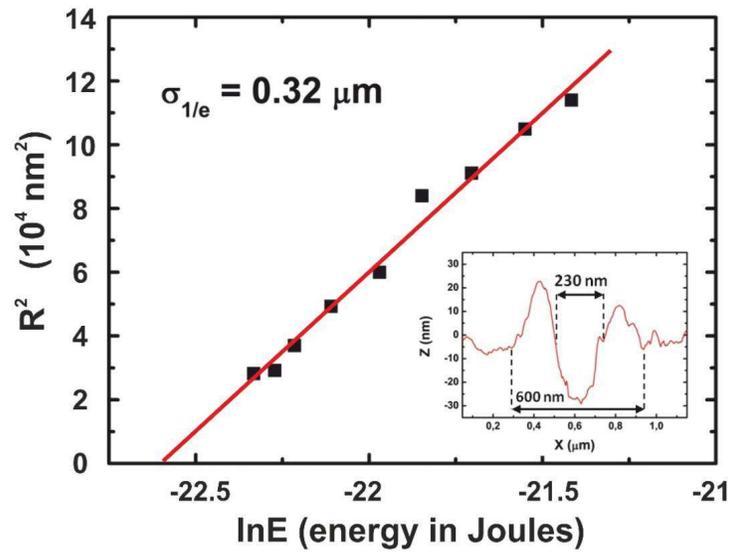}
\caption{Dependence of squared radius (R$^{2}$) of area damaged by $N \approx 2\times 10^{4}$ pulses on natural logarithm of energy per pulse (E) in Joules. The data points are fitted by linear function with a slope of 0.1024 (red line), yielding the value of Gaussian beam radius $\sigma_{1/e}$ = 0.32 $\mu$m and threshold energy of 0.16 nJ. The measurements are carried out on 30-nm Au film. Inset: AFM cross-section of the damaged area produced at $E \approx 0.38$ nJ.}\label{S1}
\end{figure*}

The method of beam size measurement~\cite{liu1982simple} is based on approximation of used focused laser beam by Gaussian spatial distribution for fluence $F(r) = (E/\pi\sigma_{1/e}^{2}) e^{(-(r/\sigma_{1/e})^{2})}$, where $\emph{r}$ is the distance from the beam center, $\sigma_{1/e}$ is the Gaussian beam size on the \emph{1/e}-level. Since the ablation threshold is a constant for given film and the number of pulses, the dependence of damaged area radii (R) on laser energy also can be fitted by Gaussian function, which is linear in $R^{2}$-lnE coordinates (Fig.~\ref{S1}).

\end{document}